\documentclass[superscriptaddress,preprint,preprintnumbers,nofootinbib,amsmath,amssymb,aps]{revtex4-2}

\usepackage{graphicx}
\usepackage{dcolumn}
\usepackage{bm}

\usepackage{booktabs}
\usepackage{tabularx}
\usepackage{lipsum}
\usepackage{amssymb}    
\usepackage{color}         
\usepackage{graphicx}     
\usepackage{mathrsfs} 
\usepackage{hyperref} 
\hypersetup{colorlinks=true,linkcolor=blue,citecolor=blue}
\usepackage{ upgreek } 
\usepackage{color,xcolor,fancybox,epsf,rotating,colordvi}

\usepackage{ulem}

\newcommand{\tanb}{\tan\!\beta}

\newcommand{\GeV}{~\rm GeV}
\newcommand{\TeV}{~\rm TeV}
\newcommand{\fbm}{{~\rm fb}^{-1}}

\begin{document}

\title{Revisiting CMSSM with Non-Universal Gaugino Masses under Current Constraints}

\author{Yabo Dong}
\email[]{dongyb@henu.edu.cn}
\affiliation{School of Physics and Electronics, Henan University, Kaifeng 475004, China}

\author{Kun Wang}
\email[Corresponding author:]{kwang@usst.edu.cn} 
\affiliation{College of Science, University of Shanghai for Science and Technology, Shanghai 200093,  China}

\author{Hailong Yuan}
\affiliation{School of Physics and Electronics, Henan University, Kaifeng 475004, China}

\author{Jingya Zhu}
\email[Corresponding author:]{zhujy@henu.edu.cn} 
\affiliation{School of Physics and Electronics, Henan University, Kaifeng 475004, China}

\author{Pengxuan Zhu}
\affiliation{ARC Centre of Excellence for Dark Matter Particle Physics, University of Adelaide, North Terrace, Adelaide SA 5005, Australia}


\date{\today}

\begin{abstract}

To address the longstanding tension between the Constrained Minimal Supersymmetric Standard Model (CMSSM) and recent experimental data, we investigate non-universal gaugino masses within an SU(5) Grand Unified Theory (GUT) framework, focusing on the $\tilde{g}$-SUGRA scenario where $\lvert M_{3} \rvert \gg \lvert M_{1} \rvert, \lvert M_{2} \rvert$. This hierarchy enables a heavier gluino, thereby evading current experimental bounds on supersymmetric particles. Our analysis reveals that precise Higgs measurements place stringent constraints on the model, requiring $\tan\beta \gtrsim 5$ and $ M_{0} \gtrsim 20 \, \tan\beta \,\text{GeV}.$ Although the $\tilde{g}$-SUGRA scenario can help reconcile the persistent $(g-2)_\mu$ anomaly, the Higgs constraints significantly restrict its parameter space, making a large contribution to $(g-2)_\mu$ challenging. We also assess the discovery prospects in upcoming dark matter direct detection experiments, including PandaX-xT (200 t.y.), LZ (projected), and XENONnT (20 t.y.), which may not fully cover the viable parameter space. In contrast, future collider experiments—such as the High-Luminosity LHC at $3\,\mathrm{ab}^{-1}$ and $\mathrm{CLIC}_{1500}$ at $2.5\,\mathrm{ab}^{-1}$—can comprehensively probe the remaining regions. These findings highlight $\tilde{g}$-SUGRA as a promising solution to the CMSSM tension and offer clear, testable predictions for upcoming collider searches.

\end{abstract}

\maketitle
\newpage

\tableofcontents


\section{\label{sec:intro}Introduction}

The Standard Model (SM) of particle physics, a foundational framework, has profoundly shaped our understanding of the fundamental constituents and forces of nature. 
Its robustness was confirmed with the discovery of the Higgs boson at the Large Hadron Collider (LHC) in 2012, a milestone that not only fulfilled a decades-long prediction, but also secured the central role of the SM in modern physics \cite{CMS:2012qbp, ATLAS:2012yve}. 
Subsequent precision measurements of the properties of the Higgs boson have consistently confirmed the predictions of the SM \cite{CMS:2022dwd, ATLAS:2022vkf}. Despite these successes, the SM has significant limitations. 
It does not adequately address critical issues such as the hierarchy problem, the anomalous magnetic moment of the muon, and the nature of dark matter, indicating the need for new theoretical developments.

The value of the muon magnetic moment predicted by the SM, $a_{\mu}^{\mathrm{SM}}$, is reported as \cite{Aoyama:2020ynm}:
\begin{align}
    a_{\mu}^{\rm SM} = 116591810(43)\times10^{-11}.
\end{align}
In 2023, Fermi National Accelerator Laboratory (FNAL) reported new measurement results for the muon magnetic moment \cite{Muong-2:2023cdq} : 
\begin{align}
    a_{\mu}^{\rm exp} = 116592059(22)\times10^{-11},
\end{align}
which are fully consistent with previous results from both Brookhaven National Laboratory (BNL) and FNAL itself \cite{Muong-2:2006rrc, Muong-2:2021ojo}. 
These data represent a 5$\sigma$ deviation from the SM predictions, highlighting a significant challenge to the existing theoretical framework. 
Further consolidating these findings, the Particle Data Group (PDG) has calculated the world average of the muon anomalous magnetic moment  as \cite{ParticleDataGroup:2024cfk} :
\begin{align}
    \Delta a_{\mu} \equiv a_{\mu}^{\rm exp} - a_{\mu}^{\rm SM} = (24.9 \pm 4.8) \times 10^{-10}.
\end{align}

In addition, the Planck Collaboration has released its final full-mission measurements of cosmological parameters \cite{Planck:2018vyg}, including the relic density of cold dark matter:
\begin{align}
    \Omega h^2 = 0.1200\pm0.0012.
\end{align}

These phenomena are difficult to explain within the framework of the SM and strongly suggest the existence of new physics beyond the Standard Model (BSM). 
To interpret such experimental results, numerous models have been proposed \cite{Chattopadhyay:1995ae, Baek:2001kca, Chattopadhyay:2001vx, Cox:2021nbo, Dermisek:2021ajd, Wang:2020xta, Wang:2020tap, Yang:2022gvz, Kawamura:2022uft, Chakraborti:2022vds, Iguro:2023tbk, Li:2023tlk, Wang:2021lwi, Wang:2020dtb}. 
Among these, low-energy supersymmetric models are most widely discussed extensions of the SM \cite{Dine:1994vc, Jungman:1995df, Maldacena:1997re, Haber:2000jh}. 

However, direct searches for SUSY particles at the LHC \cite{ATLAS:2022zwa, CMS:2022vpy} have imposed stringent constraints on the gluino and squark masses, pushing them to the multi-TeV scale. This result challenges the original low-scale SUSY paradigm. Nevertheless, electroweakinos with masses at sub-TeV level remain viable, maintaining interest in supersymmetric models from both a phenomenological and experimental perspective.

The Minimal Supersymmetric Standard Model (MSSM) \cite{Carena:1995wu, Haber:1997if, MSSMWorkingGroup:1998fiq, Giudice:2003jh} is the simplest supersymmetric extension of the SM, known for its robust ability to explain various experimental anomalies \cite{Cao:2012fz, Arbey:2012dq, Hahn:2013ria, Carena:2013qia, Beneke:2016ync, CMS:2016lcl}. 
However, it includes over 100 free parameters, making their experimental determination extremely challenging. 
In contrast, assuming the unified values for the SUSY particle masses and coupling constants at the Grand Unified Theory (GUT) scale, the Constrained MSSM (CMSSM) \cite{Dugan:1984qf, Bagger:1999rd, Ellis:2001msa, Bechtle:2013mda, Han:2016gvr, Belanger:2017vpq, GAMBIT:2017zdo} can reduces the free parameters to five.

The phenomenology of the CMSSM has been extensively studied \cite{Cao:2011sn, Bechtle:2015nua, Ellis:2012aa, Ghosh:2012dh, Nilles:1983ge, Chamseddine:1982jx, Barbieri:1982eh, Hall:1983iz, Bagnaschi:2016xfg, CMS:2015lsu, Moroi:1999zb, Izawa:1997gs, Kats:2011qh,Akula:2011aa, Akula:2012kk, Anderson:1999uia, Chamoun:2001in, Chakrabortty:2008zk,  Antusch:2009gu, Antusch:2012gv, Martin:2013aha, Kawamura:2016drh, Belyaev:2018vkl}. Among them, the gravity-mediated SUSY breaking (SUGRA) framework is highly predictive, which can naturally accommodates a 125 GeV SM-like Higgs boson \cite{Akula:2011aa, Akula:2012kk}. However, it struggles to account for the anomalous magnetic moment of the muon while predicting a 125 GeV SM-like Higgs.
Then the Non-universal gaugino masses (NUGM) extension of CMSSM is proposed to ease the tension between CMSSM and the current experimental data, which allow $M_1\neq M_2\neq M_3$ \cite{Akula:2013ioa, Aboubrahim:2021xfi, Ellis:2024ijt,  Wang:2021bcx, Wang:2018vrr}.
In Ref.~\cite{Aboubrahim:2021xfi}, a novel artificial neural network (ANN) method is used to scan the parameter space for studying the muon anomalous magnetic moment. Subsequently, the Monte Carlo method is used to investigate the capability of future colliders to cover the parameter space.
Ref.~\cite{Ellis:2024ijt} consider the scenarios involving non-universal gaugino masses, as well as non-universal slepton and Higgs masses, and then study the surviving status of benchmark points under the current experimental constraints from SUSY searches.
In Ref.~\cite{Wang:2018vrr}, a specialized mechanism based on the SU(5) Grand Unified Theory (GUT) is proposed to generate non-universal gaugino masses. The study provides a detailed discussion of dark matter constraints on this model.
Ref.~\cite{Wang:2021bcx} analyzes the 2021 Fermilab experimental data on the muon anomalous magnetic moment, offering a comprehensive discussion of potential mechanisms to reconcile the tension between the muon anomalous magnetic moment and other experimental measurements.

In this work, we revisit the CMSSM with NUGM under current constraints and focus on the scenario with $|M_3|\gg |M_1|, |M_2|$.
In such a scenario, heavy squarks and gluino can be predicted to escape the constraints on the direct search for SUSY particles. On the contrary, light electroweakinos are predicted to satisfy other experimental data. The constraints of muon anomalous magnetic moment, dark matter relic density, Higgs data, B physics, and SUSY particles direct search are considered. Additionally, we then investigate the constraints imposed by current data on this parameter space and study its surviving conditions. Then the annihilation mechanisms of dark matter are analyzed in detail.

Finally, we evaluate the prospects of future dark matter direct detection experiments and collider studies of SUSY particles to probe this scenario.

This paper is organized as follows. 
Section~\ref{sec:scenario} provides a brief overview of scanning parameter selection and the constraints considered.
Section~\ref{sec:result} presents the numerical results obtained by applying the theoretical and experimental constraints, accompanied by a comprehensive theoretical interpretation.
Finally, Section~\ref{sec:conclusions} provides a summary of our conclusions.

\section{\label{sec:scenario} Non-universal gaugino masses in $\tilde{g}$ SUGRA}

The general CMSSM has trouble both in giving a 125 GeV Higgs and in dealing with the muon anomalous magnetic moment problem under current constraints. To address the longstanding tension between the CMSSM and experimental data, the NUGM extension of the CMSSM is proposed. One of the most interesting scenarios that has arisen is $M_3 \gg M_1, M_2$, which is also known as $\tilde{g}$-SUGRA.

In this case, the gluino can significantly affect the RGEs of the squarks, contributing to their masses as the RGEs run from the GUT scale to the electroweak scale. Conversely, at the one-loop level, the gluino does not enter into the RGEs of sleptons, leaving their masses unaffected. As a result, heavy squarks and gluino are predicted to evade the constraints on the direct search for SUSY particles, while light electroweakinos are predicted to satisfy other experimental data. In fact, the light smuon can provide a significant contribution to the g-2 to match the experimental result. Such a scenario predicts heavy squarks and gluino, allowing them to evade current experimental constraints on their masses. At the same time, the presence of relatively light sleptons can lead to a number of interesting phenomenological consequences.

The scenario with $M_3 \gg M_1, M_2$ are considered in this work. We use the \textsf{SuSpect-2.52} \cite{Djouadi:2002ze} package to implement the theoretical and experimental constraints in our analysis, assuming $\text{sign}(\mu) = +1$. The remaining parameters are selected within the following ranges:
\begin{equation}
\begin{aligned}
 M_0,\ |M_1|,\ |M_2| <&\ 1\  \TeV,\\
  1\  \TeV < |M_3| <&\ 10\  \TeV,\\
 |A_0| <&\ 10\  \TeV,\\ 
 1 < \tanb <&\ 50.
 \end{aligned}
 \end{equation}

The following constraints are considered in our analysis:
\begin{itemize}	
	\item[(1)] The lightest CP-even Higgs boson is assumed to be the SM-like Higgs boson, with a mass of $125\pm2 \ \GeV$ \cite{CMS:2012qbp, ATLAS:2012yve, ATLAS:2015egz, CMS:2014fzn}. 
	\item[(2)] The constraints on the squark and gluino masses are derived from the latest results of the LHC searches \cite{CMS:2017gbz}:
	\begin{equation}
	m_{\tilde{g}}>2\ \mathrm{TeV},\quad m_{\tilde{t}}>0.7\ \mathrm{TeV},\quad m_{\tilde{q}_{1,2}}>2\ \mathrm{TeV},
	\end{equation}
	while the chargino and slepton mass constraints are derived from LEP results \cite{ALEPH:2005ab}:
	\begin{equation}m_{\tilde{\tau}}>93.2\ \mathrm{GeV},\quad m_{\tilde{\chi}^{\pm}}>103.5\ \mathrm{GeV}.\end{equation}
	\item[(3)] Direct searches for low-mass and high-mass resonances have been conducted at LEP, Tevatron, and LHC. These constraints are implemented in our analysis using the \textsf{HiggsSignal-2.6.0} \cite{Bechtle:2020uwn} and \textsf{HiggsBounds-5.10.0} \cite{Bechtle:2020pkv} packages.
	\item[(4)]The dark matter relic density constraints are derived from the Planck observations \cite{Planck:2018vyg}, while the direct search results for dark matter are based on the 2022 findings of the LUX-ZEPLIN (LZ) experiment \cite{LZ:2022lsv}. The relic density and the dark matter annihilation cross-section are calculated using the \textsf{micrOMEGAs-5.3.41} package \cite{Belanger:2001fz, Belanger:2010pz}. Since the relic density may also include contributions from other dark matter species, only the upper bound is applied in this analysis, given as $0 < \Omega h^2 < 0.12$\,.
	\item[(5)] The constraints from B physics, such as $B\rightarrow s\gamma$, $B_s\rightarrow \mu^+\mu^-$, and $B^+\rightarrow \tau^+\nu$ are adopted based on the latest results provided by the PDG 2024 \cite{ParticleDataGroup:2024cfk}:
	 \begin{equation}
	 \begin{aligned}
	Br&(B\rightarrow s\gamma) = (3.49 \pm 0.38) \times 10^{-4},\\
        Br&(B^+\rightarrow \tau^+\nu) = (1.09\pm 0.48) \times 10^{-4},\\ 
        Br&(B_s\rightarrow \mu^+\mu^-) = (3.01 \pm 0.87) \times 10^{-9}.        
	 \end{aligned}
	\end{equation}
    \item[(6)]The constraints on the muon anomalous magnetic moment ($\Delta a_{\mu}$), defined as the difference between the experimental measurement and the SM prediction, are characterized by the following features \cite{Muong-2:2023cdq, ParticleDataGroup:2024cfk}
    \begin{align}
	\Delta a_{\mu} \equiv a_{\mu}^{\mathrm{exp}} - a_{\mu}^{\mathrm{SM}} = (24.9 \pm 4.8) \times 10^{-10}. 
    \end{align}
    The muon anomalous magnetic moment is calculated at the two-loop level using the \textsf{GM2Calc-2.1.0} package \cite{Athron:2015rva, Athron:2021evk}.
    \item[(7)] The constraints from direct searches for supersymmetric particles at LEP, Tevatron, and LHC are incorporated into the analysis. The \textsf{SModelS-3.0.0} package \cite{Kraml:2013mwa, Ambrogi:2017neo, Alguero:2021dig} is used to calculate the expected signal strength $r$, which is used to evaluate the consistency of the model with experimental results.
\end{itemize}

\section{\label{sec:result}Results and discussion} 

\begin{figure*}[!tbp] 
\centering 
\includegraphics[width=1\linewidth]{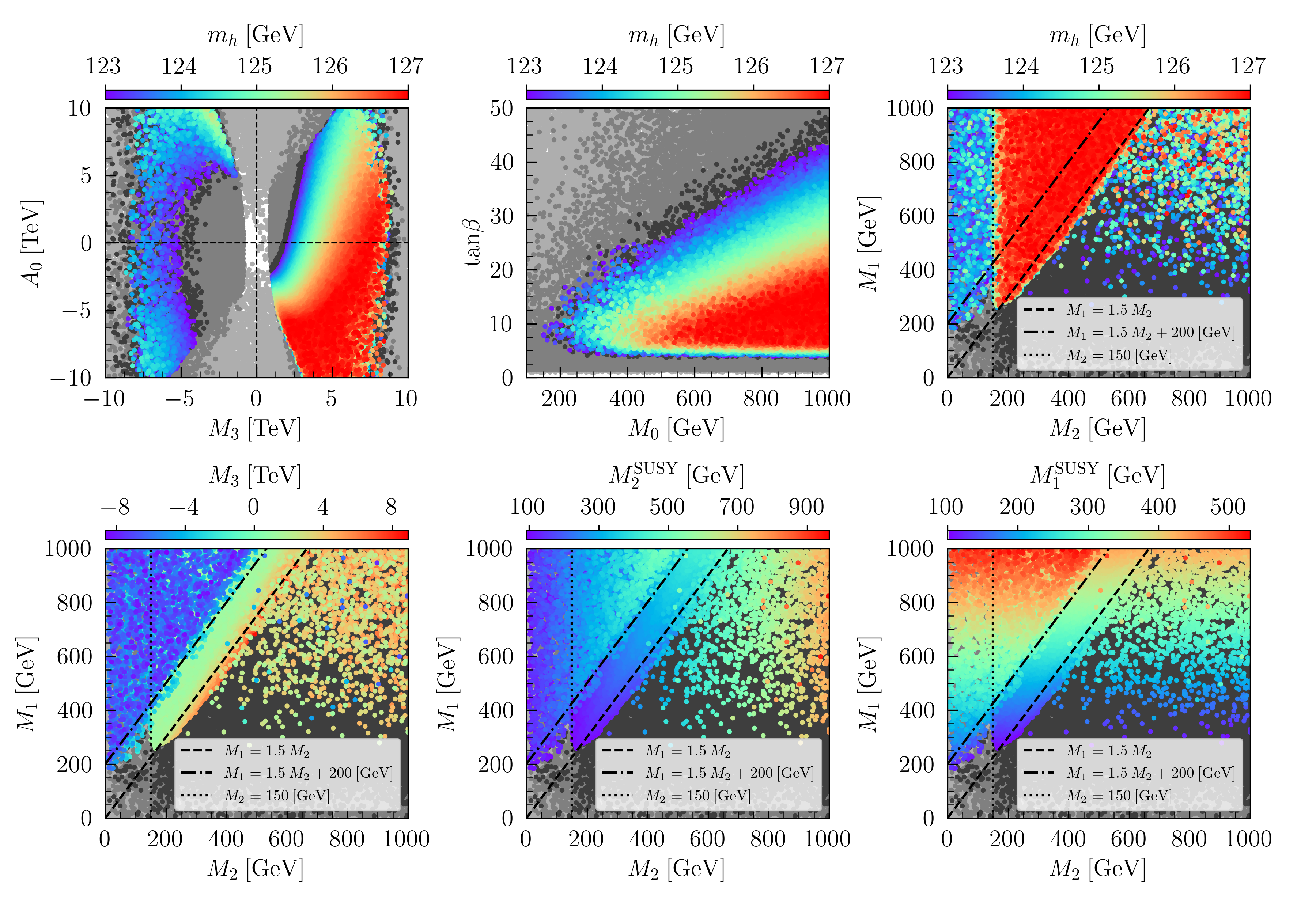}
\caption{\label{f01} 
    Surviving samples in the $A_0$ versus $M_3$ plane (upper left), the $\tanb$ versus $M_0$ plane (upper middle), and the $M_1$ versus $M_2$ plane (upper right and lower three), with colors indicating the SM-like Higgs mass $m_h$ (upper three), the value of input parameter $M_3$ (lower left), the value of $M_2$ at the soft SUSY scale (lower middle), and the value of $M_1$ at the soft SUSY scale (lower right), respectively. The light gray samples are excluded by theoretical constraint, the gray samples are excluded that cannot give a suitable SM-like Higgs mass, and the black samples are excluded by the dark matter data. The samples with a larger value of $m_h$ are projected on top of smaller ones (upper panels) and the samples with a smaller value of $M_3$ are projected on top of larger ones (lower panels).}
\end{figure*}

In Fig.~\ref{f01}, we present all surviving samples in the $A_0$ versus $M_3$ plane (upper left), the $\tanb$ versus $M_0$ plane (upper middle), and the $M_1$ versus $M_2$ plane (upper right and lower three), respectively. 
Their colors indicate the SM-like Higgs mass $m_h$ (upper three), the value of input parameter $M_3$ (lower left), the value of $M_2$ at the soft SUSY scale (lower middle), and the value of $M_1$ at the soft SUSY scale (lower right), respectively. 
The light gray samples are excluded by the theoretical constraints, the gray samples are excluded that cannot give a suitable SM-like Higgs mass, and the black samples are excluded by the dark matter data. 
The samples with a larger value of $m_h$ are projected on top of smaller ones (upper panels), and the samples with a smaller value of $M_3$ are projected on top of larger ones (lower panels).
Several conclusions can be drawn from these figures:
\begin{itemize}
    \item The primary constraints on $M_3$, $A_0$, $M_0$, and $\tanb$ arise from theoretical considerations and Higgs data.  In contrast, the primary constraints on $M_1$ and $M_2$ are imposed by dark matter data. The reasons for these differences will be discussed in more detail later.

    \item In the upper right and lower three panels of Fig.~\ref{f01}, the surviving samples in the  $M_1$ versus $M_2$ plane can be broadly classified into three categories:
\begin{itemize}	
	\item $\textbf{Class\  A}$: $M_3>0$ GeV, $M_1/M_2 \gtrsim 1.5$\,.
	\item $\textbf{Class\  B}$: $M_3>0$ GeV, $M_1/M_2 \lesssim 1.5$\,.
	\item $\textbf{Class\  C}$: $M_3<0$ GeV, $M_1 \gtrsim 1.5 M_2 + 200$ GeV.
\end{itemize}   
\end{itemize}
Due to the use of more accurate calculation tools and the consideration of newer experimental constraints, the survival status of parameters in our study differs significantly from that in Ref.~\cite{Wang:2018vrr}. Specifically, samples with larger values of $\tan \beta$ ($\tan \beta \gtrsim 42$ or $\tan \beta \gtrsim 25$ with $M_0 \approx 400\ \text{GeV}$) are excluded in our analysis, whereas they are allowed in Ref.~\cite{Wang:2018vrr}. 
\begin{figure*}[!tbp] 
\centering 
\includegraphics[width=1\linewidth]{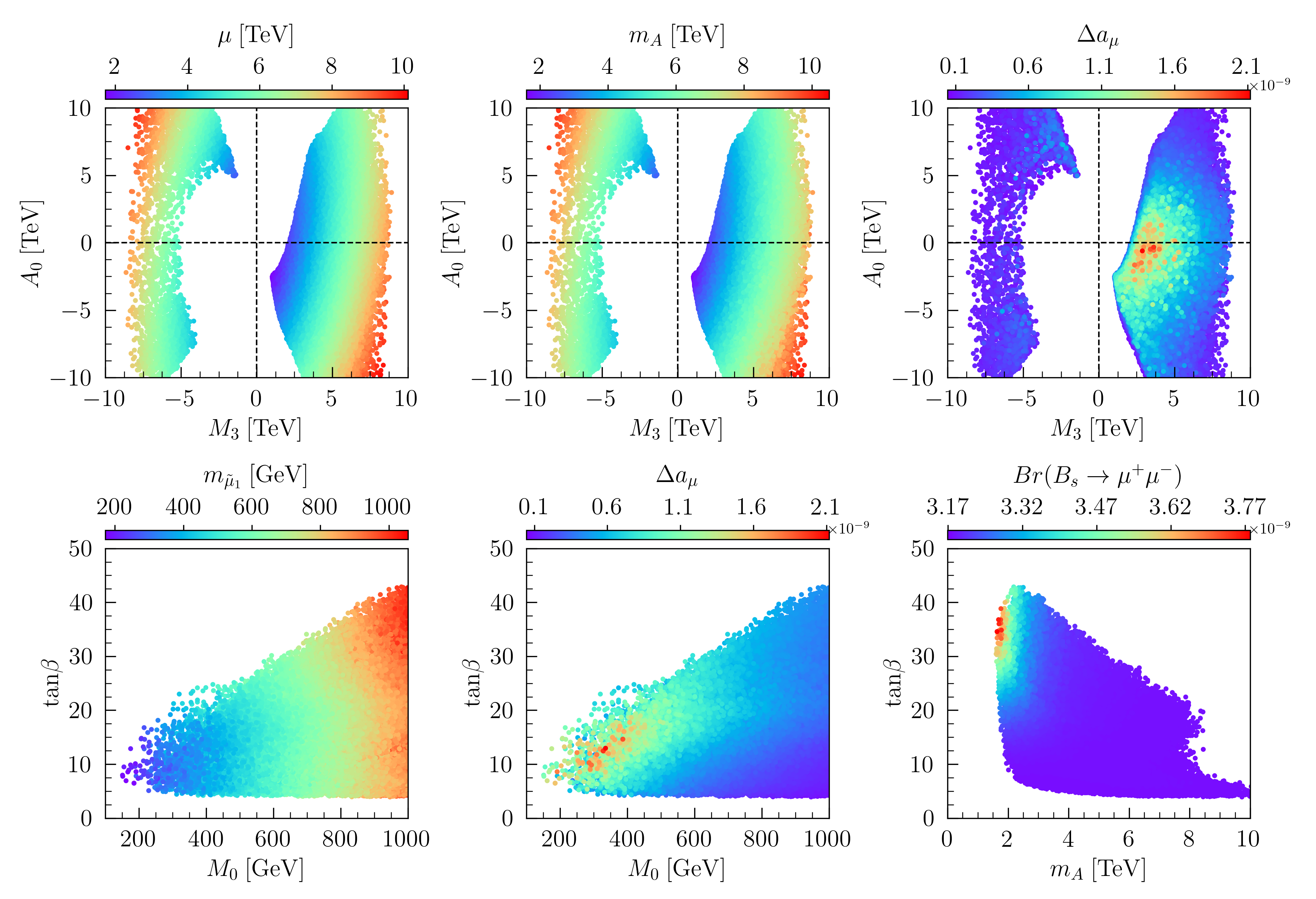}
\caption{\label{f02} 
    Surviving samples in the $A_0$ versus $M_0$ planes (upper three), the $\tanb$ versus $M_0$ planes (lower left and lower middle), and the $\tanb$ versus $m_A$ plane (lower right), with colors indicating the higgsino mass parameter $\mu$ (upper left), the pseudoscalar Higgs mass $m_A$ (upper middle), the SUSY contributions to muon g-2 $\Delta a_{\mu}$ (upper right and lower middle), the lighter smuon mass $m_{\tilde{\mu}_1}$ (lower left), the branching ratio of $B_s \to \mu^+\ \mu^-$ (Br($B_s \to \mu^+\ \mu^-$)) (lower right), respectively. For the upper right and lower three panels, the samples with larger values of colors are projected on top of smaller ones.} 
\end{figure*}

In Fig.~\ref{f02}, surviving samples are shown in the $A_0$ versus $M_0$ planes (upper three), the $\tanb$ versus $M_0$ planes (lower left and lower middle), and the $\tanb$ versus $m_A$ plane (lower right), with colors indicating the higgsino mass parameter $\mu$ (upper left), the pseudoscalar particle Higgs mass $m_A$ (upper middle), the SUSY contributions to muon g-2 $\Delta a_{\mu}$ (upper right and lower middle), the lighter smuon mass $m_{\tilde{\mu}_1}$ (lower left), the branching ratio of $B_s \to \mu^+\ \mu^-$ (Br($B_s \to \mu^+\ \mu^-$)) (lower right), respectively. 
Meanwhile, samples with larger values of colors are projected on top of smaller ones for the upper right and lower three panels.
By combining Fig.~\ref{f01} and Fig.~\ref{f02}, the following conclusions can be drawn:
\begin{itemize}
    \item From the upper panels of Fig.~\ref{f01}, it is evident that Higgs data impose a strong constraint on the parameter space. Specifically, the data requires  $\tanb\gtrsim5$ and $M_0\gtrsim20 \tanb$ GeV. In addition, a negative $M_3$ is favor to predict a light Higgs mass, whereas a positive $M_3$ is favor to predict a heavy Higgs mass. 
    
    \item From the lower left panel of Fig.~\ref{f01}, it can be observed that samples with negative $M_3$ can only surviving in $\textbf{Class\  C}$. In other words, to obtain a negative value of $M_3$ requires $M_1 \gtrsim 1.5 M_2 + 200$ GeV. In contrast, the samples with positive $M_3$ only need $M_2\gtrsim200$ GeV.  
    
    \item From the lower middle and right panels of Fig.~\ref{f01}, it can be observed that, for the same $M_1$ ($M_2$) at the GUT scale, the samples in $\textbf{Class\  C}$ result in larger $M_1$ ($M_2$) values at the soft SUSY scale compared to samples in other classes after RGEs running. This is because a negative $M_3$ provides a positive contribution to $M_1$ ($M_2$) during the RGEs evolution, whereas a positive $M_3$ leads to a negative contribution. These results are consistent with Eqs.(38) and (39) in the Appendix A in Ref.~\cite{Wang:2018vrr}.

    \item From the upper right panel of Fig.~\ref{f01}, it can be observed that samples in $\textbf{Class\  A}$ tend to predict a heavier SM-like Higgs. Furthermore, a heavy SM-like Higgs is also favored by moderate values of $\tanb$ (around 10) and relatively large $M_0$ (larger than 400 GeV), as shown in the upper middle panel of Fig.~\ref{f01}. 
    
    \item By combining the upper left and middle panels of Fig.~\ref{f02}, it can be observed that the Higgsino mass  $\mu$ is nearly equal to the pseudoscalar Higgs mass  $m_A$, and both are directly proportional to $M_3$. 
    This relationship arises because, in supersymmetric models, the parameter $\mu$ takes the form:
    \begin{align}\label{mu}
        \mu^2=\frac{M_{H_d}^2-M_{H_u}^2\tan^2\beta}{\tan^2\beta-1}-\frac{m_Z^2}{2}.
    \end{align}
    Since $\tanb \gg 1$, we have $\tan^2\beta - 1 \approx \tan^2\beta$ and $|M_{H_u}| \gg |M_{H_d}|$. Under these approximations, the term $M_{H_d}^2 - M_{H_u}^2 \tan^2\beta$ reduces to $M_{H_u}^2 \tan^2\beta$, and the first term in Eq.~\eqref{mu} simplifies to approximately $M_{H_u}^2$. Furthermore, since $|M_{H_u}| \gg |m_Z|$, it follows that $\mu \approx M_{H_u}$. 
    At the same time, the pseudoscalar Higgs mass parameter $m_A$ is given by:
    \begin{align}
        m_A^2 = \frac{\tan^2\beta + 1}{\tan^2\beta - 1} \left( M_{H_d}^2 - M_{H_u}^2 \right) - m_Z^2.
    \end{align}
    With $\tanb \gg 1$ and $|M_{H_u}| \gg |M_{H_d}|$, one can find that $\tan^2\beta - 1 \approx \tan^2\beta + 1$ and $M_{H_d}^2 - M_{H_u}^2 \approx M_{H_u}^2$. This leads to $m_A \approx M_{H_u} \approx \mu$. 
    According to the Eqs.(59) in the Appendix A in Ref.~\cite{Wang:2018vrr}, $M_{H_u}$ is nearly directly proportional to $M_3$. Consequently, both $\mu$ and $m_A$ are approximately directly proportional to $M_3$.

    \item By combining the upper right and lower middle panels of Fig.~\ref{f02}, it appears that small values of $|A_0|$, $M_3$, and $M_0$ tend to provide a larger contribution to $\Delta a_{\mu}$. In particular, a sample with $M_0 \sim 250 \, \mathrm{GeV}$, $|A_0| \lesssim 1 \, \mathrm{TeV}$, $M_3 \sim 4 \, \mathrm{TeV}$, and $\tanb \sim 10$ yields a sizable contribution to $\Delta a_{\mu}$ ($\sim 2.1 \times 10^{-9}$). This result is consistent with Ref.\cite{Ellis:2024ijt}, but differs from Ref.\cite{Aboubrahim:2021xfi}, as the latter considers a different set of theoretical assumptions and constraints. Our analysis incorporates additional constraints, leading to slight variations in the allowed parameter space.
    In SUSY, the primary contribution to $\Delta a_{\mu}$ comes from smuon interactions and is given by \cite{Moroi:1995yh, Endo:2013lva}:
    \begin{align}
    a_{\mu}^{\tilde{\mu}} \approx a_0 \frac{1 + \delta^{\text{2-loop}}}{1 + \Delta_{\mu}} \left( \frac{\tanb \cdot (100 \, \mathrm{GeV})^2}{m_{\tilde{\mu}_L}^2 m_{\tilde{\mu}_R}^2 / (M_1 \mu)} \right) \left( \frac{f_N}{1/6} \right),
    \end{align}
    where $a_0 = 1.5 \times 10^{-10}$. 
    This expression shows that the smuon contribution is proportional to $\tanb$ and inversely proportional to $m_{\tilde{\mu}}^4$. Therefore, achieving a significant supersymmetric contribution to $g-2$ requires light sleptons and large $\tanb$. 
    Smaller $M_0$ values can produce lighter smuons, as seen in the lower left panel of Fig.~\ref{f02}. 
    However, scenarios with small $M_0$ and large $\tanb$ encounter significant theoretical constraints, including issues with RGEs due to the presence of non-perturbative effects or Landau poles.
    Additionally, these scenarios are also constrained by Higgs data, as they struggle to produce an SM-like Higgs with a mass of $125 \, \mathrm{GeV}$. Consequently, achieving a extremely large contribution to $g-2$ in this framework is highly challenging.

    \item From the lower right panel of Fig.~\ref{f02}, it can be observed that a lighter $m_A$ ( $\sim$ 2 TeV) and a larger $\tanb$ ($\sim$ 35) result in a higher branching ratio for $B_s \to \mu^+ \mu^-$. This behavior can be explained within SUSY models, where the branching ratio of $B_s \to \mu^+ \mu^-$ is given by:
    \begin{align}
        \text{Br}(B_s \to \mu^+ \mu^-) \propto \frac{m_t^4 \mu^2 A_t^2 \tan^6\beta}{m_A^4 m_t^4}.
    \end{align}
    From this expression, it is evident that $\text{Br}(B_s \to \mu^+ \mu^-)$ is directly proportional to $\tan^6\beta$ and inversely proportional to $m_A^4$. This explains why a larger $\tanb$ and a smaller $m_A$ enhance the branching ratio.

\end{itemize}

\begin{figure*}[!tbp] 
\centering 
\includegraphics[width=1\linewidth]{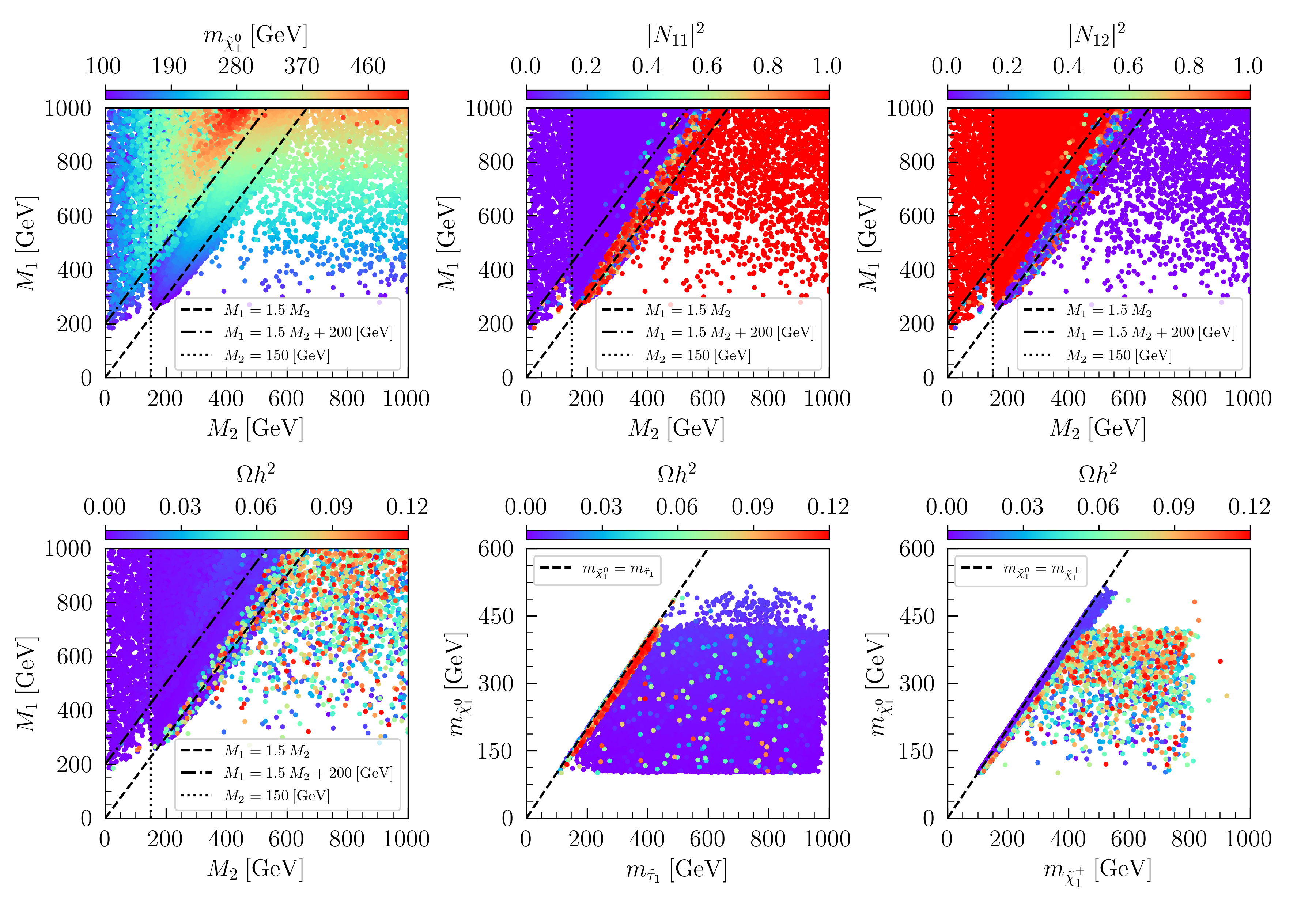}
\caption{\label{f03} 
    Surviving samples in the $M_1$ versus $M_2$ planes (upper three and lower left), the lightest neutralino mass $m_{\tilde{\chi}_1^0}$ versus lighter stau mass $m_{\tilde{\tau}_1}$ plane (lower middle), and the lightest neutralino mass $m_{\tilde{\chi}_1^0}$ versus the lightest chargino mass $m_{\tilde{\chi}_1^{\pm}}$ plane (lower right), with colors indicating the lightest neutralino mass $m_{\tilde{\chi}_1^0}$ (upper left), the bino component in lightest neutralino (upper middle), the wino component in lightest neutralino (upper right), and the dark matter relic density $\Omega h^2$ (lower three), respectively. The samples with smaller values of $M_3$ are projected on top of larger ones for the upper left panels, the samples with larger values of $|N_{11}|^2$ are projected on top of smaller ones for the upper middle panel, the samples with smaller value of $|N_{12}|^2$ are projected on top of larger ones for the upper right panel, and the samples with larger value of $\Omega h^2$ are projected on top of smaller ones for the lower three panels.} 
\end{figure*}

In Fig.~\ref{f03}, to understand the dark matter annihilation mechanisms of the surviving samples, we project them onto the $M_1$ versus $M_2$ planes (upper three and lower left), the lightest neutralino mass $m_{\tilde{\chi}_1^0}$ versus the lighter stau mass $m_{\tilde{\tau}_1}$ plane (lower middle), and the lightest neutralino mass $m_{\tilde{\chi}_1^0}$ versus the lightest chargino mass $m_{\tilde{\chi}_1^{\pm}}$ plane (lower right).
The colors indicate the lightest neutralino mass $m_{\tilde{\chi}_1^0}$ (upper left), the bino component of the lightest neutralino (upper middle), the wino component of the lightest neutralino (upper right), and the dark matter relic density $\Omega h^2$ (lower three), respectively. 
Meanwhile, samples with smaller values of $M_3$ are projected on top of larger ones for the upper left panels, and the samples with larger values of $|N_{11}|^2$ are projected on top of smaller ones for the upper middle panel, the samples with a smaller value of $|N_{12}|^2$ are projected on top of larger ones for the upper right panel, and the samples with larger value of $\Omega h^2$ are projected on top of smaller ones for the lower three panels.
The following conclusions can be drawn:
\begin{itemize}
    
    \item From the upper left panel of Fig.~\ref{f03}, it is evident that the lightest neutralino mass $m_{\tilde{\chi}_1^0}$ is nearly proportional to $M_2$ for samples with $M_1/M_2 \gtrsim 1.5$, while it is nearly proportional to $M_1$ for samples with $M_1/M_2 \lesssim 1.5$. This behavior may suggests the dominant component of the lightest neutralino $\tilde{\chi}_1^0$, as $M_1$ and $M_2$ are the mass parameters associated with the bino and wino, respectively. 
    Furthermore, samples with $M_3<0$ predict a heavier $\tilde{\chi}_1^0$ compared to samples with $M_3>0$ for the same $M_2$. This is attributed to the additional contribution of $M_3$ to $M_2$ at the soft SUSY scale.

    \item From the upper middle and right panels of Fig.~\ref{f03}, it can be observed that the lightest neutralino $\tilde{\chi}_1^0$ is predominantly wino-like when $M_1/M_2 \gtrsim 1.5$, while it is predominantly bino-like when $M_1/M_2 \lesssim 1.5$. Additionally, $\tilde{\chi}_1^0$ generally does not exhibit wino-bino mixing, except for samples with $M_1/M_2 \approx 1.5$.

    \item By combining the upper middle, upper right, and lower left panels of Fig.~\ref{f03}, it becomes clear that a larger dark matter relic density $\Omega h^2$ is only achieved for samples with $M_1/M_2 \lesssim 1.5$, where $\tilde{\chi}_1^0$ is predominantly bino-like or wino-bino mixing. However, the $\Omega h^2$ predicted for them is often excessively large and strongly constrained by experimental dark matter data. This limitation explains the scarcity of surviving samples with $M_1/M_2 \lesssim 1.5$. 
    In contrast, samples with $M_1/M_2 \gtrsim 1.5$, where $\tilde{\chi}_1^0$ is predominantly wino-like, tend to predict a small $\Omega h^2$ to escape the dark matter relic density data.

    \item From the lower left and middle panels of Fig.~\ref{f03}, it can be observed that the mass of the lightest neutralino $m_{\tilde{\chi}^0_1}$ and the mass of the lighter smuon $m_{\tilde{\tau}_1}$ are nearly degenerate when $M_1/M_2 \lesssim 1.5$. In contrast, $m_{\tilde{\tau}_1}$ becomes significantly larger than $m_{\tilde{\chi}^0_1}$ when $M_1/M_2 \gtrsim 1.5$. This behavior may provide insight into the annihilation mechanisms of $\tilde{\chi}^0_1$. 
    When $\tilde{\chi}^0_1$ is bino-like, its annihilation may be primarily mediated by interactions with $\tilde{\tau}_1$. Since $\tilde{\tau}_1$ is only slightly heavier than $\tilde{\chi}^0_1$, the annihilation process occurs relatively slowly, allowing for the accumulation of a sufficient relic density.

    \item From the lower left and right panels of Fig.~\ref{f03}, it can be observed that when $M_1/M_2 \gtrsim 1.5$, the mass of the lightest neutralino $m_{\tilde{\chi}^0_1}$ nearly degenerate with the mass of the lightest chargino $m_{\tilde{\chi}_1^{\pm}}$, which aligns with theoretical expectations for wino-like samples. Consequently, $\tilde{\chi}^0_1$ can consistently co-annihilate with $\tilde{\chi}_1^{\pm}$ for these samples. 
    However, due to the high annihilation rate associated with this co-annihilation process, it fails to produce a sufficient relic density ($\Omega h^2$). 
    In contrast, when $\tilde{\chi}_1^0$ is bino-like, the mass of $\tilde{\chi}_1^{\pm}$ is significantly larger than that of $\tilde{\chi}_1^0$, making co-annihilation between the two particles inefficient.
    
\end{itemize}

\begin{figure*}[!tbp]
\centering    
\begin{minipage}[b]{\linewidth}
\centering
\includegraphics[width=0.99\linewidth]{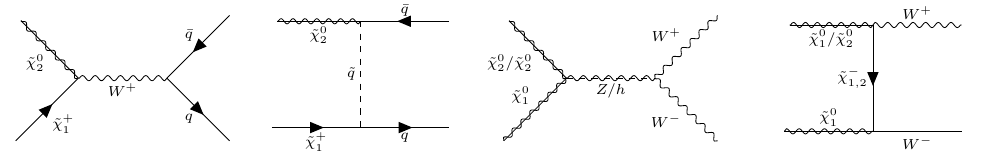} 
\vspace{20pt} 
\end{minipage}    
\begin{minipage}[t]{\linewidth}
\centering
\includegraphics[width=0.99\linewidth]{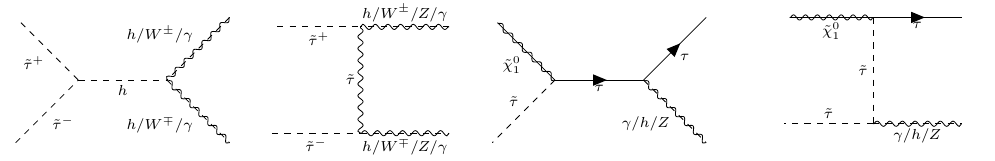}
\end{minipage}    
\caption{Main Feynman diagrams illustrating $\tilde{\chi}_{1}^{\pm}$ coannihilation (upper panels) and $\tilde{\tau}, \tilde{\chi}^0_1$ hybrid annihilation (lower panels).\label{fd}}
\end{figure*}

After examining the annihilation mechanisms of $\tilde{\chi}^0_1$ for the surviving samples, we identify the following primary mechanisms. 
The main Feynman diagrams illustrating these dominant annihilation mechanisms are shown in Fig.~\ref{fd}. 
\begin{itemize}
    \item Stau annihilation : $\tilde{\tau}_{1}^{+}\tilde{\tau}_{1}^{-}\rightarrow hh, W^+W^-, t\bar{t}, \tilde{\tau}_{1}^{+}\tilde{\tau}_{1}^{-}\rightarrow\tau^{+}\tau^{-}$. 

    \item Neutralino-stau coannihilation  : $\tilde{\chi}_{1}^{0}\tilde{\tau}_{1}^{\pm}\to\tau^{\pm}h/A$.

    \item Neutralino annihilation  : $\tilde{\chi}_{1}^{0}\tilde{\chi}_{1,2}^{0}\rightarrow W^+W^-$.
    \item Neutralino-chargino coannihilation  : $\tilde{\chi}_{1,2}^{0}\tilde{\chi}_{1}^{\pm} \rightarrow u\bar{d}, c\bar{s}, t\bar{b}$.

    \item Chargino annihilation  : $\tilde{\chi}_{1}^{+}\tilde{\chi}_{1}^{-}\to W^{+}W^{-}$.
    
\end{itemize}

The surviving samples can be categorized into four classes based on their dominant annihilation mechanisms:

\begin{equation}
\begin{aligned}
\tilde{\tau}\tilde{\chi}^0_1\ \mathrm{coannihilation}:&\ \frac{m_{\tilde{\tau}_1}}{m_{\tilde{\chi}^0_1}} < 1.1\,.\\
\tilde{\chi}^{0}_1\tilde{\chi}^{\pm}_1\ \mathrm{coannihilation}:&\ \frac{m_{\tilde{\tau}_1}}{m_{\tilde{\chi}^0_1}} \gtrsim 1.1\,,\ 1< \frac{m_{\tilde{\chi}^{\pm}_1}}{m_{\tilde{\chi}^0_1}} \lesssim 1.05\,. \\
\tilde{\chi}^{\pm}_1\tilde{\chi}^0_1\ \mathrm{coannihilation}:&\ \frac{m_{\tilde{\tau}_1}}{m_{\tilde{\chi}^0_1}} \gtrsim 1.1\,,\ 1.05\lesssim\frac{m_{\tilde{\chi}^{\pm}_1}}{m_{\tilde{\chi}^0_1}} < 1.1\,. \\
\tilde{\chi}^{0}_1\ \mathrm{annihilation}:&\ \frac{m_{\tilde{\tau}_1}}{m_{\tilde{\chi}^0_1}} \gtrsim 1.1\,,\ \frac{m_{\tilde{\chi}^{\pm}_1}}{m_{\tilde{\chi}^0_1}} \approx 1.1\, .
\end{aligned}
\end{equation}


The dominant annihilation channels for these scenarios are as follows:
\begin{itemize}

    \item $\tilde{\tau}\tilde{\chi}^0_1$ coannihilation: the primary channels are stau annihilation and neutralino-stau coannihilation.
    \item $\tilde{\chi}^{0}_1\tilde{\chi}^{\pm}_1$ coannihilation: the major channels are neutralino-chargino coannihilation and neutralino annihilation.
    \item $\tilde{\chi}^{\pm}_1\tilde{\chi}^0_1$ coannihilation: the main channels are chargino annihilation and neutralino-chargino coannihilation.
    \item $\tilde{\chi}^{0}_1$ annihilation: the dominant channel is neutralino annihilation.
\end{itemize}

\begin{table*}[!tbp]
	\centering
	\caption{\label{t02}The top 10 annihilation channels and their relative contributions to $\sigma v$ for the four benchmark points.}
        \resizebox*{1\textwidth}{!}{
	\begin{tabular}{>{\hspace{15pt}}r<{\hspace{15pt}}>{\hspace{15pt}}r<{\hspace{15pt}}>{\hspace{15pt}}r<{\hspace{15pt}}>{\hspace{15pt}}r<{\hspace{15pt}}}
		\toprule
		  P1 & P2 & P3 & P4 \\
		\midrule
		$\tilde{\chi}_1^+  \tilde{\chi}_2^0  \rightarrow u \bar{d}, 0.08 $ & 
		$\tilde{\tau}_1^+ {\tilde{\tau}_1^-} \rightarrow h h , 0.17$&
		$\tilde{\chi}_1^0 \tilde{\chi}_1^0 \rightarrow W^+ W^-,0.18$&
            $\tilde{\chi}_1^0  \tilde{\tau}_1 \rightarrow \gamma \tau ,0.23$\\  	
		$\tilde{\chi}_1^+  \tilde{\chi}_2^0  \rightarrow c \bar{s}, 0.08 $&
		$\tilde{\tau}_1^+ \tilde{\tau}_1^- \rightarrow  W^+ W^-, 0.16$&
		$\tilde{\chi}_1^0 \tilde{\chi}_2^0 \rightarrow W^+ W^-,0.12$&
            $\tilde{\tau}_1^+  \tilde{\tau}_1^- \rightarrow \tau^+ \tau^- ,0.21$\\
		$\tilde{\chi}_1^+  \tilde{\chi}_2^0  \rightarrow t \bar{b}, 0.07 $&
		$\tilde{\chi}_1^0  \tilde{\tau}_1^{\pm}  \rightarrow  h\tau^{\pm},0.12$&
		$\tilde{\chi}_1^+ \tilde{\chi}_1^0 \rightarrow u \bar{d},0.07$&
            $\tilde{\chi}_1^0  \tilde{\tau}_1^{\pm} \rightarrow Z \tau^{\pm} ,0.09$\\
		$\tilde{\chi}_2^0 \tilde{\chi}_2^0 \rightarrow W^+ W^-,0.07$ &
		$\tilde{\tau}_1^+ \tilde{\tau}_1^- \rightarrow  t \bar{t}, 0.11$&
		$\tilde{\chi}_1^+ \tilde{\chi}_1^0 \rightarrow c \bar{s},0.07$ &
            $\tilde{\chi}_1^0 \tilde{\chi}_1^+ \rightarrow \tau h,0.08$ \\
		$\tilde{\chi}_1^0 \tilde{\chi}_2^0 \rightarrow W^+ W^-,0.07$ & 
		$\tilde{\chi}_1^0 \tilde{\tau}_1^{\pm}  \rightarrow \gamma \tau^{\pm} ,0.10$&
		$\tilde{\chi}_2^0 \tilde{\chi}_2^0 \rightarrow W^+ W^-,0.06$ &
            $\tilde{\tau}_1^+  \tilde{\tau}_1^- \rightarrow h h,0.07$\\
		$\tilde{\chi}_1^0  \tilde{\chi}_2^0  \rightarrow Z W^+ ,0.06$ &
		$\tilde{\tau}_1^+  \tilde{\tau}_1^-  \rightarrow Z Z ,0.08$ &
		$\tilde{\chi}_1^+ \tilde{\chi}_2^0  \rightarrow u \bar{d} , 0.06$ &             
            $\tilde{\tau}_1^+  \tilde{\tau}_1^-  \rightarrow W^+ W^-,0.06$\\
		$\tilde{\chi}_1^+  \tilde{\chi}_1^+  \rightarrow W^+ W^+,0.05$ &
		$\tilde{\tau}_1^+  \tilde{\tau}_1^-   \rightarrow  \tau^+ \tau-, 0.08$&
		$\tilde{\chi}_1^+  \tilde{\chi}_2^0  \rightarrow  c \bar{s} , 0.06$ &
            $\tilde{\tau}_1^+  \tilde{\tau}_1^- \rightarrow \gamma \gamma,0.06$\\
		$\tilde{\chi}_1^+  \tilde{\chi}_1^0   \rightarrow u \bar{d} ,0.04$ &
		$\tilde{\chi}_1^0   \tilde{\tau}_1^{\pm}   \rightarrow  Z \tau^{\pm}, 0.06$&
		$\tilde{\chi}_1^+ \tilde{\chi}_1^0  \rightarrow t \bar{b} , 0.05$ &
            $\tilde{\tau}_1^+  \tilde{\tau}_1^- \rightarrow t \bar{t},0.05$\\
		$\tilde{\chi}_1^+ \tilde{\chi}_1^0  \rightarrow c \bar{s}  ,0.04$ &
		$\tilde{\chi}_1^0 \tilde{\tau}_1^{\pm} \rightarrow W^{\pm} \nu_{\tau}, 0.06 $&
		$\tilde{\chi}_1^+  \tilde{\chi}_2^0  \rightarrow  t \bar{b} ,0.04$& 
            $\tilde{\chi}_1^0 \tilde{\tau}_1^{\pm} \rightarrow W^{\pm} \nu_{\tau},0.04$\\
		$\tilde{\chi}_1^0 \tilde{\chi}_1^0  \rightarrow W^+ W^-,0.04$ &
		$\tilde{\tau}_1^+  \tilde{\tau}_1^- \rightarrow \gamma \gamma, 0.03 $   &
		$\tilde{\chi}_1^+ \tilde{\chi}_2^0 \rightarrow Z W^+,0.03$& 
            $\tilde{\tau}_1^+  \tilde{\tau}_1^- \rightarrow Z Z,0.03$\\		
		\bottomrule
	\end{tabular}}
\end{table*}

\begin{table*}[!tbp]
	\centering
	\caption{\label{t03}Detailed information for the four surviving benchmark samples.}
        \resizebox*{1\textwidth}{!}{
	\begin{tabular}{>{\hspace{30pt}}l<{\hspace{30pt}}>{\hspace{30pt}}c<{\hspace{30pt}}>{\hspace{30pt}}c<{\hspace{30pt}}>{\hspace{30pt}}c<{\hspace{30pt}}>{\hspace{30pt}}c<{\hspace{30pt}}}
		\toprule
		  & P1 & P2 & P3 & P4 \\
		\midrule
		$\tan \beta$          &12.63 & 9.59 & 5.81 & 12.88 \\
	$A_0 \mathrm{~[GeV]} $ &-4540.0& -9491.1& 3812.3 & -623.3  \\
	$M_0  \mathrm{~[GeV]}$ &582.3 & 790.1 & 990.2 & 216.9  \\
	$M_1  \mathrm{~[GeV]}$ & 595.0 &786.2 & 303.7 & 646.8\\
	$M_2  \mathrm{~[GeV]}$ & 347.8 & 828.5 & 99.0 & 945.0  \\
        $M_3  \mathrm{~[GeV]}$ & 1586.7 & -6091.1 & -5387.8 & 2276.6  \\
	$\mu  \mathrm{~[GeV]}$ &2803.7 & 5850.1 & 6165.0 & 2581.0 \\
	$m_h  \mathrm{~[GeV]}$ &125.6 & 123.8 & 123.2 & 123.1 \\ 
	$m_A  \mathrm{~[GeV]}$ & 2779.4 & 5915.4 & 6315.2 & 2619.0\\
	$m_{\Tilde{t}_1}  \mathrm{~[GeV]}$ &1814.0 & 8551.1  & 7412.2 & 3318.0\\
	$m_{\Tilde{\tau}_1}  \mathrm{~[GeV]}$ &397.3 & 416.0 & 897.1 &  273.6\\
	$m_{\Tilde{\mu}_1}  \mathrm{~[GeV]}$  & 616.9 & 805.4 & 916.6 & 314.1\\
	$m_{\Tilde{\nu}_e}  \mathrm{~[GeV]}$ &614.1 & 883.0 & 913.1  & 605.1\\
	$m_{\Tilde{\nu}_{\tau}}  \mathrm{~[GeV]}$ &530.3 & 747.5 & 905.3  & 605.1\\
	$m_{\Tilde{\chi}_1^{\pm}} \mathrm{~[GeV]}$  &272.2 & 865.6 & 193.1  & 759.1\\
	$Br(B_s \rightarrow {\mu}^+ {\mu}^-)[10^{-9}]$  & 3.19 & 3.18 &3.18 & 3.18\\
	$Br(B \rightarrow {\tau} {\nu})[10^{-4}]$ &0.837 & 0.837 & 0.837 & 0.836 \\
	$Br(B \rightarrow s \gamma)[10^{-4}]$ &3.330 & 3.354 & 3.355 & 3.377\\
	$\Delta a_{\mu}^{SUSY}[10^{-10}]$ & 3.98 & 1.90 & 0.61 & 6.49\\
	$m_{\Tilde{\chi}_1^0} \mathrm{~[GeV]}$ & 248.4 & 406.0 & 170.9 & 267.1\\
	$m_{\Tilde{\chi}_2^0} \mathrm{~[GeV]}$ &270.4 & 813.1 & 189.8 & 716.2\\
	$\sigma_{\text{SI}}$ [$10^{-48}$ cm$^{-1}$]  & 10.7 & 1.03 & 5.13 & 5.34\\
	$\sigma_P^{SD}$[$10^{-46}$ cm$^{-1}$] &25.1 & 0.61 & 2.04 & 13.3\\
	$\sigma_N^{SD}$ [$10^{-46}$ cm$^{-1}$] &50.2 & 0.835 & 2.39  & 21.3\\
	$\Omega h^2 $  & 0.109 & 0.114 & 0.118 & 0.113\\
		\bottomrule
	\end{tabular}}
\end{table*}

We provide detailed annihilation information for four benchmark points, including their top 10 annihilation channels and the relative contributions to $\sigma v$, in Table~\ref{t02}. Additional information about these benchmark points is summarized in Table~\ref{t03}. In contrast to Refs.~\cite{Aboubrahim:2021xfi, Ellis:2024ijt, Wang:2021bcx}, which focus primarily on parameter space constraints, our study includes a comprehensive analysis of the dark matter annihilation mechanisms, providing a more detailed understanding of their phenomenological implications.

\begin{figure*}[!tbp] 
	\centering 
	\includegraphics[width=1\linewidth]{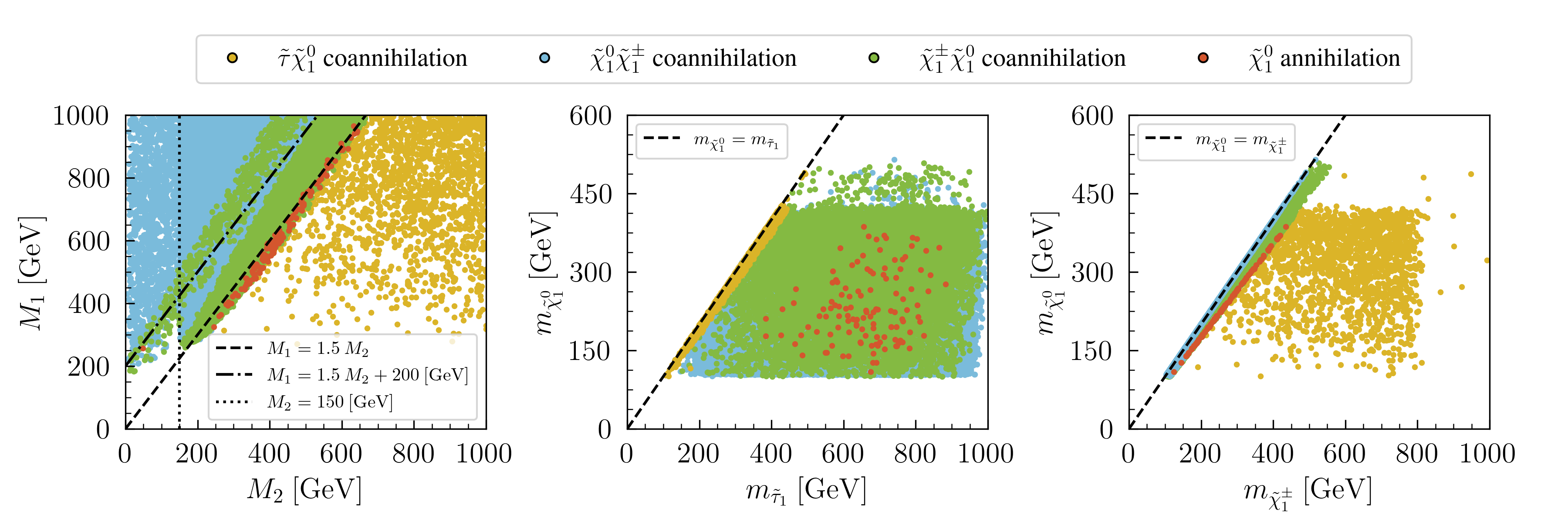}
	\caption{\label{f04} 
	Surviving samples in the $M_1$ versus $M_2$ planes (left), the lightest neutralino mass $m_{\tilde{\chi}_1^0}$ versus lighter stau mass $m_{\tilde{\tau}_1}$ plane (middle), and the lightest neutralino mass $m_{\tilde{\chi}_1^0}$ versus the lightest chargino mass $m_{\tilde{\chi}_1^{\pm}}$ plane(right). The main annihilation mechanisms for the samples are indicated by their colors: yellow samples correspond to $\tilde{\tau}\tilde{\chi}^0_1$ coannihilation, blue samples represent $\tilde{\chi}^{\pm}_1\tilde{\chi}^0_1$ coannihilation, green samples correspond to $\tilde{\chi}^0_1\tilde{\chi}^{\pm}_1$ coannihilation, and red samples indicate $\tilde{\chi}^0_1$ annihilation.} 
\end{figure*}

In Fig.~\ref{f04}, surviving samples are shown in the $M_1$ versus $M_2$ plane (left), the lightest neutralino mass $m_{\tilde{\chi}_1^0}$ versus the lighter stau mass $m_{\tilde{\tau}_1}$ plane (middle), and the lightest neutralino mass $m_{\tilde{\chi}_1^0}$ versus the lightest chargino mass $m_{\tilde{\chi}_1^{\pm}}$ plane (right).
The main annihilation mechanisms for the samples are indicated by their colors: yellow samples correspond to $\tilde{\tau}\tilde{\chi}^0_1$ coannihilation, blue samples represent $\tilde{\chi}^{\pm}_1\tilde{\chi}^0_1$ coannihilation, green samples correspond to $\tilde{\chi}^0_1\tilde{\chi}^{\pm}_1$ coannihilation, and red samples indicate $\tilde{\chi}^0_1$ annihilation.
The following conclusions can be drawn:
\begin{itemize}
    \item By comparing the upper middle and right panels of Fig.~\ref{f03}, along with the left panel of Fig.~\ref{f04}, one can see that the dominant annihilation mechanisms of the samples with $M_1/M_2 \gtrsim 1.5$ are $\tilde{\chi}^{0}_1\tilde{\chi}^{\pm}_1$ coannihilation, which reflects the tendency of wino-like $\tilde{\chi}^0_1$ to engage in this mechanism. In contrast, for the bino-like samples with $M_1/M_2 \lesssim 1.5$, $\tilde{\tau}\tilde{\chi}^0_1$ coannihilation is the main mechanism. The dominant mechanism can also be $\tilde{\chi}^{\pm}_1\tilde{\chi}^0_1$ coannihilation and $\tilde{\chi}^{0}_1$ annihilation when $M_1/M_2 \approx 1.5$ or $M_1 \approx 1.5M_2 +200\ \mathrm{GeV}$, depending on the relative masses $\tilde{\chi}^0_1$ and $\tilde{\chi}^{\pm}_1$.
    
    \item From the middle and right panels of Fig.~\ref{f04}, it is observed that for samples with degenerate masses of $\tilde{\tau}$ and $\tilde{\chi}^0_1$, the main annihilation mechanism is $\tilde{\tau}\tilde{\chi}^0_1$ coannihilation.
    In contrast, for samples with non-degenerate masses of $\tilde{\tau}$ and $\tilde{\chi}^0_1$, annihilation mechanisms include $\tilde{\chi}^{\pm}_1\tilde{\chi}^0_1$ coannihilation, $\tilde{\chi}^{0}_1\tilde{\chi}^{\pm}_1$ coannihilation, and $\tilde{\chi}^{0}_1$ annihilation. The particular annihilation mechanism depends on the mass relation between $\tilde{\chi}^{0}_1$ and $\tilde{\chi}^{\pm}_1$. Specifically, the main annihilation mechanism is $\tilde{\chi}^{\pm}_1\tilde{\chi}^0_1$ coannihilation or $\tilde{\chi}^{0}_1$ annihilation when $m_{\tilde{\chi}^{\pm}_1}/m_{\tilde{\chi}^0_1}\approx1$ or $m_{\tilde{\chi}^{\pm}_1}/m_{\tilde{\chi}^0_1}\approx 1.1$. And the main annihilation mechanisms can be $\tilde{\chi}^{0}_1\tilde{\chi}^{\pm}_1$ coannihilation when $1\lesssim m_{\tilde{\chi}^{\pm}_1}/m_{\tilde{\chi}^0_1}\lesssim1.1$.
\end{itemize}

\begin{figure*}[!tbp] 
\centering 
\includegraphics[width=0.8\linewidth]{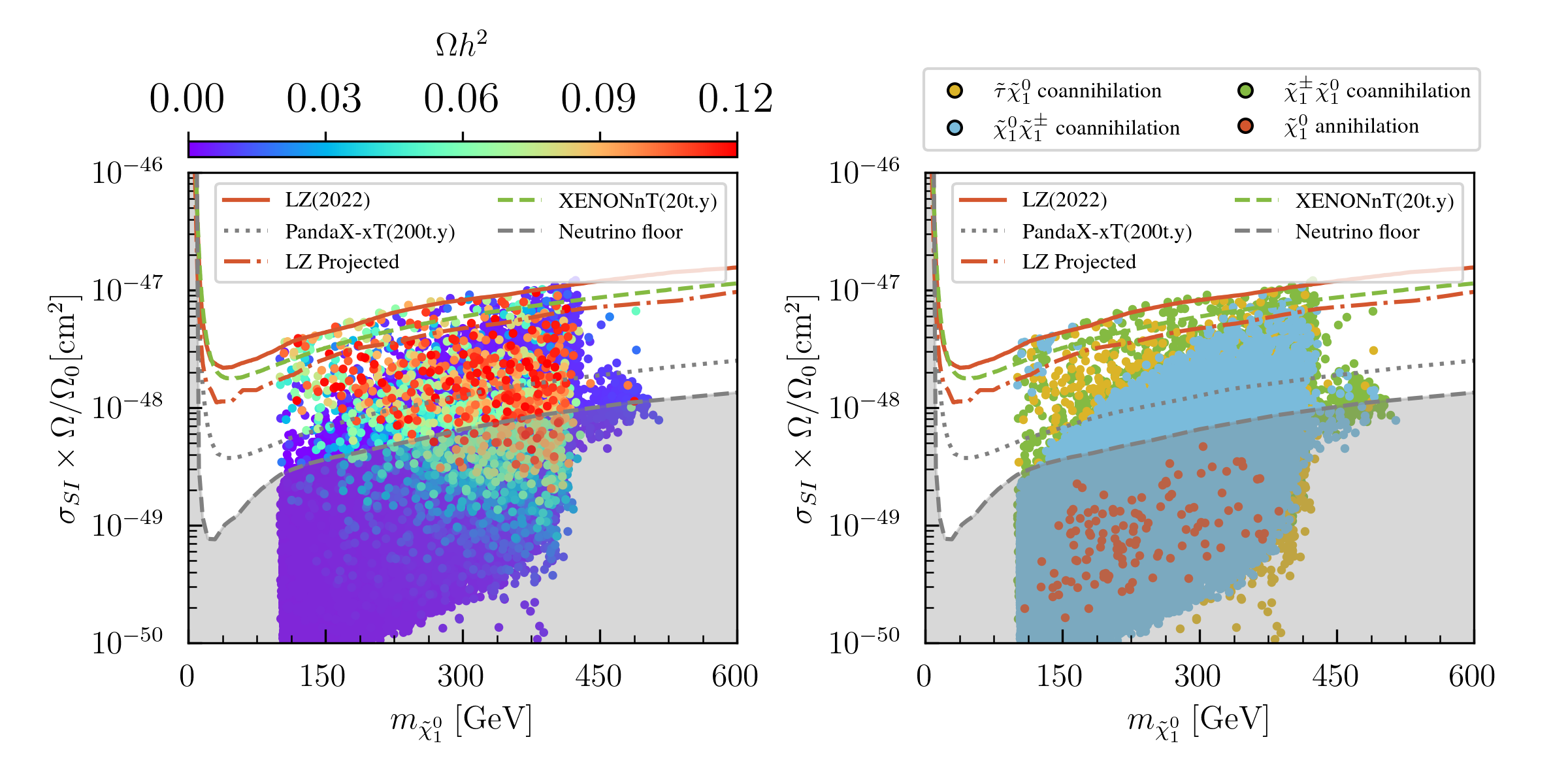}
\caption{\label{f05} 
    Surviving samples in the rescaled spin independence (SI) DM-nucleon cross-section $\sigma_{SI}$ (rescaled by $\Omega/\Omega _0$) versus $m_{\tilde{\chi}^0_1}$ panels. For the left panel, colors indicate dark matter relic density $\Omega h^2$. 
    For the right panel, the main annihilation mechanisms of yellow samples are $\tilde{\tau}\tilde{\chi}^0_1$ coannihilation, the main annihilation mechanisms of blue samples are $\tilde{\chi}^{\pm}_1\tilde{\chi}^0_1$ coannihilation, the main annihilation mechanisms of green samples are $\tilde{\chi}^{0}_1\tilde{\chi}^{\pm}_1$ coannihilation, and the main annihilation mechanisms of red samples are $\tilde{\chi}^{0}_1$ annihilation. 
    The red solid curves indicate the spin-independent (SI) DM-nucleon cross-section detection limits of LZ (2022) \cite{LZ:2022lsv}. 
    The black dotted, red dot-dashed, and green dashed curves indicate the future detection limits PandaX-xT(200t.y) \cite{PandaX:2024oxq}, LZ Projected \cite{LZ:2018qzl}, and XENONnT(20t.y) \cite{XENON:2020kmp}, respectively. 
    The gray shaded region indicated the neutrino floor \cite{Billard:2013qya}.
    } 
\end{figure*}

\begin{figure*}[!tbp] 
\centering 
\includegraphics[width=1\linewidth]{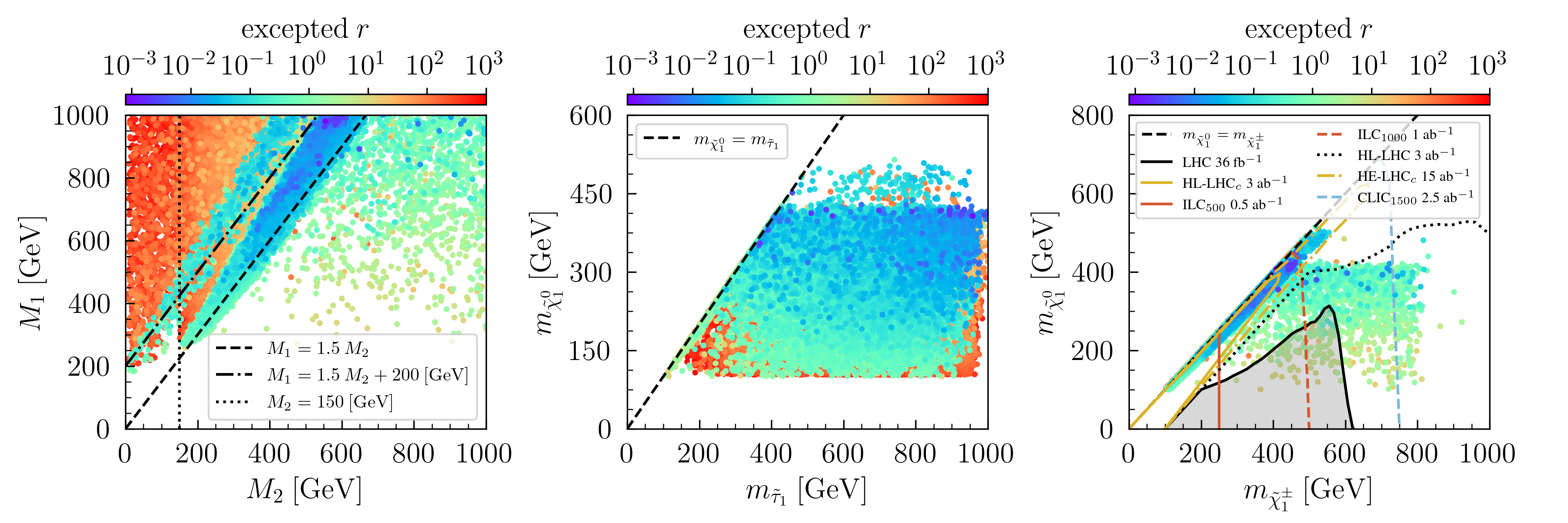}
\caption{\label{f06} 
    Surviving samples in the $M_1$ versus $M_2$ planes (left), the lightest neturalino mass $m_{\tilde{\chi}_1^0}$ versus the lighter stau mass $m_{\tilde{\tau}_1}$ planes (middle) and  $m_{\tilde{\chi}_1^0}$ versus the lighter chargino mass $m_{\tilde{\chi}_1^{\pm}}$ planes (righ) with Colors indicate the expected $r$ of signal strength.
    The black, yellow, and red solid curves of the right panel indicate the detection limits at LHC with $36\fbm$ \cite{ATLAS:2017mjy, CMS:2016eep}, HL-LHC compressed (HL-LHC$_c$) with 3 $\mathrm{ab}^{-1}$ \cite{TheATLAScollaboration:2014nwe}, and 500 GeV International Linear Collider (ILC) with 0.5 $\mathrm{ab}^{-1}$ \cite{Berggren:2020tle}, respectively. The red dashed, black dotted, yellow dot-dashed, and green dashed curves indicate the future detectors of 1000 GeV ILC with 1 $\mathrm{ab}^{-1}$ \cite{Berggren:2020tle}, HL-LHC with 3 $\mathrm{ab}^{-1}$ \cite{TheATLAScollaboration:2014nwe}, HE-LHC compressd (HE-LHC$_c$) with 15 $\mathrm{ab}^{-1}$ \cite{CidVidal:2018eel} and 1500 GeV Compact Linear Collider (CLIC$_{1500}$) with 2.5 $\mathrm{ab}^{-1}$ \cite{CLIC:2018fvx}. The samples with smaller values of excepted $r$ are projected on top of larger ones.
 }
\end{figure*}

In Fig.~\ref{f05}, the surviving samples are shown in the rescaled spin-independent (SI) DM-nucleon cross-section $\sigma_{SI}$ (rescaled by $\Omega/\Omega_0$) versus $m_{\tilde{\chi}^0_1}$. In the left panel, the colors represent the dark matter relic density $\Omega h^2$. In the right panel, the colors indicate the primary annihilation mechanisms: yellow for $\tilde{\tau}\tilde{\chi}^0_1$ coannihilation, blue for $\tilde{\chi}^{\pm}_1\tilde{\chi}^0_1$ coannihilation, green for $\tilde{\chi}^0_1\tilde{\chi}^{\pm}_1$ coannihilation, and red for $\tilde{\chi}^0_1$ annihilation. 
The red solid curve shows the spin-independent DM-nucleon cross-section detection limits from LZ in 2022 \cite{LZ:2022lsv}. The black dotted, red dot-dashed, and green dashed curves represent future detection limits from PandaX-xT (200t.y) \cite{PandaX:2024oxq}, the projected LZ \cite{LZ:2018qzl}, and XENONnT (20t.y) \cite{XENON:2020kmp}, respectively. The gray shaded region indicates the neutrino floor \cite{Billard:2013qya}.
The following conclusions can be drawn: 
\begin{itemize}
    \item All surviving samples predict a smaller rescaled $\sigma_{SI}$, with most falling below the detection limits of LZ 2022. From the left panel of Fig.~\ref{f06}, it can be seen that samples with larger $\Omega h^2$ tend to predict larger rescaled $\sigma_{SI}$, while those dominated by $\tilde{\chi}^{0}_1$ annihilation predict very small rescaled $\sigma_{SI}$, all of which fall below the neutrino floor.  
    \item By combining Fig.~\ref{f05} with Fig. 4 in Ref.~\cite{Wang:2024ozr} and Fig. 5 in Ref.~\cite{Zhao:2022pnv}, one can observe that, compared to the higgsino-like dark matter, the bion-like dark matter tends to predict a significantly smaller rescaled $\sigma_{SI}$.

    \item Even with future dark matter detection experiments such as PandaX-xT (200t.y), the LZ Projected, and XENONnT (20t.y), it will remain difficult to cover all surviving samples. Many samples lie below the neutrino floor, making them nearly undetectable by dark matter experiments and requiring examination in collider experiments.
\end{itemize}

In Fig.~\ref{f06}, the surviving samples are presented in the $M_1$ versus $M_2$ planes (left), the lightest neutralino mass $m_{\tilde{\chi}_1^0}$ versus the lighter stau mass $m_{\tilde{\tau}_1}$ planes (middle) and $m_{\tilde{\chi}_1^0}$ versus the lighter chargino mass $m_{\tilde{\chi}_1^{\pm}}$ planes (right). 
The colors indicate the expected signal strength ratio $r$, calculated by \textsf{SModelS-3.0.0} \cite{Kraml:2013mwa, Ambrogi:2017neo, Alguero:2021dig}. 
The right panel highlights the coverage of current and future collider experiments, including the LHC, High-Luminosity LHC (HL-LHC), High-Energy LHC (HE-LHC), the International Linear Collider (ILC), and the Compact Linear Collider (CLIC)~\cite{EuropeanStrategyforParticlePhysicsPreparatoryGroup:2019qin}. 
Specifically, the black, yellow, and red solid curves of the right panel indicate the detection limits at LHC with $36\fbm$ \cite{ATLAS:2017mjy, CMS:2016eep}, HL-LHC compressed (HL-LHC$_c$) with 3 $\mathrm{ab}^{-1}$ \cite{TheATLAScollaboration:2014nwe}, and 500 GeV International Linear Collider (ILC) with 0.5 $\mathrm{ab}^{-1}$ \cite{Berggren:2020tle}, respectively. The red dashed, black dotted, yellow dot-dashed, and green dashed curves indicate the future detectors of 1000 GeV ILC with 1 $\mathrm{ab}^{-1}$ \cite{Berggren:2020tle}, HL-LHC with 3 $\mathrm{ab}^{-1}$ \cite{TheATLAScollaboration:2014nwe}, HE-LHC compressed (HE-LHC$_c$) with 15 $\mathrm{ab}^{-1}$ \cite{CidVidal:2018eel} and 1500 GeV Compact Linear Collider (CLIC$_{1500}$) with 2.5 $\mathrm{ab}^{-1}$ \cite{CLIC:2018fvx}.
Moreover, samples with smaller values of the expected $r$ are projected on top of those with larger values.
One can draw several conclusions from Fig.~\ref{f06}:
\begin{itemize}

    
    \item Since an expected $r$ of signal strength greater than 1 is considered excluded by the experiment, it can be observed from the left panel of Fig.~\ref{f06} that, compared to the samples with $M_1/M_2\lesssim1.5$, the samples with $M_1/M_2\gtrsim1.5$ are more likely to be excluded by experiments. There are exceptions for the sample with $M_3>0$ and $M_1/M_2\approx1.5$ or the samples with $M_3<0$ and $M_1\approx1.5 M_2 +200$ GeV. These samples are also hard to be excluded by current SUSY particles direct search experiment.

    \item By combining the left panel of Fig.~\ref{f04} with the left and middle panels of Fig.~\ref{f06}, an interesting observation can be made. Samples that are more easily excluded by current SUSY particles search experiments correspond to $\tilde{\chi}^{\pm}_1\tilde{\chi}^{0}_1$ coannihilation. Conversely, when the annihilation mechanism of the sample is $\tilde{\tau}\tilde{\chi}^0_1$ coannihilation, $\tilde{\chi}^{0}_1\tilde{\chi}^{\pm}_1$ coannihilation, or $\tilde{\chi}^{0}_1$ annihilation, the experimental constraints are much weaker. These observations indicate that the direct searches for SUSY particles experiment impose strong constraints on samples with non-degenerate mass of $\tilde{\tau}$ and $\tilde{\chi}^0_1$ ($m_{\tilde{\tau}_1}/m_{\tilde{\chi}^0_1} \gtrsim 1.1$) and degenerate mass of $\tilde{\chi}^{\pm}_1$ and $\tilde{\chi}^0_1$ ($m_{\tilde{\chi}^{\pm}_1}/m_{\tilde{\chi}^0_1}\approx1$).

    \item From the right panel of Fig.~\ref{f06}, it can be observed that hadron colliders, including the LHC, HL-LHC, and HE-LHC, have very limited constraints on the compressed mass interval of the samples. Even with increased energy and luminosity, it remains challenging to cover all of these samples. In contrast, lepton colliders, such as the ILC and CLIC, demonstrate significantly better coverage in this region. They can effectively probe almost all samples with masses less than half of their collision energy, even with relatively modest integral luminosities. 
    In general, when the HL-LHC achieves an integral luminosity of $3 \, \mathrm{ab}^{-1}$ and the CLIC$_{1500}$ reaches $2.5 \, \mathrm{ab}^{-1}$, all surviving samples will be fully covered by these experiments.

\end{itemize}

In this paper, we qualitatively discuss the sensitivity of future colliders, such as HL-LHC, HE-LHC, ILC, and CLIC, in probing the surviving parameter space based on experimental exclusion limits. In contrast, Ref.\cite{Aboubrahim:2021xfi} uses Monte Carlo methods for a quantitative assessment of the HL-LHC and HE-LHC reach in different search channels, while Ref.\cite{Ellis:2024ijt} investigates the survival of benchmark points under current SUSY search constraints. Furthermore, the phenomenology of $(g-2)_\mu$, dark matter, and $B$ physics for specific benchmark points, such as $M_1=M_2$ and $M_1=5M_2$, has been discussed in detail in Ref.~\cite{Wang:2021bcx}. In contrast, our work extends this analysis by exploring scenarios with arbitrary ratios between $M_1$ and $M_2$.
Additionally, we conduct a detailed study of dark matter annihilation mechanisms and their implications for direct detection and collider experiments. Compared to our previous work in Ref.~\cite{Wang:2018vrr}, this study employs more sophisticated computational tools for parameter space analysis and includes a thorough discussion of the ability to cover the parameter space of future collider experiments.

\section{Conclusions}
\label{sec:conclusions}

We investigated non-universal gaugino masses within an SU(5) GUT framework to reconcile the CMSSM with current experimental data. In particular, we focused on the $\tilde{g}$-SUGRA scenario, where $|M_3| \gg |M_1|, |M_2|$, so that the gluino mass significantly exceeds the masses of the other gauginos. Our analysis shows that precise Higgs data restricts the parameter space by requiring $\tan\beta \gtrsim 5$ and $M_0 \gtrsim 20\,\tan\beta\,\mathrm{GeV}$. Moreover, negative $M_3$ typically favors a lighter Higgs boson, whereas positive $M_3$ tends to yield a heavier one.

A substantial supersymmetric contribution to the muon anomalous magnetic moment $(g-2)_\mu$ demands both light sleptons and large $\tan\beta$, yet combining small $M_0$ with high $\tan\beta$ often makes it challenging to achieve a 125\,GeV Higgs. Nevertheless, this framework can still accommodate the muon $(g-2)$ anomaly. We also observed that negative $M_3$ raises $M_1^\mathrm{SUSY}$ or $M_2^\mathrm{SUSY}$ through renormalization-group running, whereas positive $M_3$ lowers them. Depending on the ratio of $M_1$ to $M_2$, the lightest neutralino $\tilde{\chi}_1^0$ can be wino-like, bino-like, or a wino-bino mixture; only bino-like states typically yield a larger dark matter relic density.

Concerning dark matter annihilation, the primary channels include stau annihilation, neutralino--stau coannihilation, neutralino--chargino coannihilation, and neutralino--neutralino annihilation. While these processes efficiently reduce the dark matter relic abundance, many samples predict extremely small spin-independent scattering cross sections $\sigma_{SI}$, sometimes below the neutrino floor, making them difficult to probe via direct detection experiments. On the other hand, future collider searches hold strong prospects for testing this scenario: once the High-Luminosity LHC reaches $3\,\mathrm{ab}^{-1}$ and CLIC$_{1500}$ achieves $2.5\,\mathrm{ab}^{-1}$, the entire parameter space—including the wino-like region—can be thoroughly examined. Our results thus indicate that the $\tilde{g}$-SUGRA scenario provides a consistent resolution to the CMSSM tension, offering clear, testable predictions for collider experiments.

\begin{acknowledgments}
Yabo Dong thanks Rui Zhu for helpful discussion. This work was supported by the National Natural Science Foundation of China under Grant No. 12275066 and by the startup research funds of Henan University. 
\end{acknowledgments}


\bibliographystyle{apsrev4-1}
\bibliography{apssamp}

\end{document}